\documentclass[sigconf,screen,authorversion]{acmart} 

\usepackage[nameinlink]{cleveref}  

\usepackage{caption}
\usepackage{subcaption}
\usepackage{tikz}
\usetikzlibrary{arrows,automata}

\copyrightyear{2022}
\acmYear{2022}
\setcopyright{acmlicensed}
\acmConference[SIGCSE 2022]{Proceedings of the 53rd ACM Technical Symposium on Computer Science Education V. 1}{March 3--5, 2022}{Providence, RI, USA}
\acmBooktitle{Proceedings of the 53rd ACM Technical Symposium on Computer Science Education V. 1 (SIGCSE 2022), March 3--5, 2022, Providence, RI, USA}
\acmPrice{15.00}
\acmDOI{10.1145/3478431.3499420}
\acmISBN{978-1-4503-9070-5/22/03}

\newcommand{\circlednumber}[1]{\raisebox{.5pt}{\textcircled{\raisebox{-.9pt} {#1}}}}

\usepackage{csquotes}

\usepackage{comment}
\usepackage{pgfplots}
\usepgfplotslibrary{colorbrewer}
\usetikzlibrary{pgfplots.statistics, pgfplots.colorbrewer}
\usepackage{pgfplotstable}

\pgfplotsset{compat=1.16}
\pgfplotsset{
    box plot/.style={
        /pgfplots/.cd,
        only marks,
        mark=-,
        font=\sffamily,
        mark size=1em,
        /pgfplots/error bars/.cd,
        y dir=plus,
        y explicit,
    },
    box plot box/.style={
        /pgfplots/error bars/draw error bar/.code 2 args={%
            \draw  ##1 -- ++(1em,0pt) |- ##2 -- ++(-1em,0pt) |- ##1 -- cycle;
        },
        /pgfplots/table/.cd,
        y index=2,
        y error expr={\thisrowno{3}-\thisrowno{2}},
        /pgfplots/box plot
    },
    box plot top whisker/.style={
        /pgfplots/error bars/draw error bar/.code 2 args={%
            \pgfkeysgetvalue{/pgfplots/error bars/error mark}%
            {\pgfplotserrorbarsmark}%
            \pgfkeysgetvalue{/pgfplots/error bars/error mark options}%
            {\pgfplotserrorbarsmarkopts}%
            \path ##1 -- ##2;
        },
        /pgfplots/table/.cd,
        y index=4,
        y error expr={\thisrowno{2}-\thisrowno{4}},
        /pgfplots/box plot
    },
    box plot bottom whisker/.style={
        /pgfplots/error bars/draw error bar/.code 2 args={%
            \pgfkeysgetvalue{/pgfplots/error bars/error mark}%
            {\pgfplotserrorbarsmark}%
            \pgfkeysgetvalue{/pgfplots/error bars/error mark options}%
            {\pgfplotserrorbarsmarkopts}%
            \path ##1 -- ##2;
        },
        /pgfplots/table/.cd,
        y index=5,
        y error expr={\thisrowno{3}-\thisrowno{5}},
        /pgfplots/box plot
    },
    box plot median/.style={
        /pgfplots/box plot,
    },
}

\begin{document}

\fancyhead{} 

\title{Preventing Cheating in Hands-on Lab Assignments}

\author{Jan Vykopal}
\orcid{0000-0002-3425-0951}
\affiliation{
  \institution{Masaryk University}
  \country{Czech Republic}
}
\email{vykopal@ics.muni.cz}

\author{Valdemar Švábenský}
\orcid{0000-0001-8546-280X}
\affiliation{
  \institution{Masaryk University}
  \country{Czech Republic}
}
\email{svabensky@ics.muni.cz}

\author{Pavel Seda}
\orcid{0000-0002-6689-1980}
\affiliation{
  \institution{Masaryk University}
  \country{Czech Republic}
}
\email{seda@fi.muni.cz}

\author{Pavel Čeleda}
\orcid{0000-0002-3338-2856}
\affiliation{
  \institution{Masaryk University}
  \country{Czech Republic}
}
\email{celeda@ics.muni.cz}

\begin{abstract} 
Networking, operating systems, and cybersecurity skills are exercised best in an authentic environment. Students work with real systems and tools in a lab environment and complete assigned tasks. Since all students typically receive the same assignment, they can consult their approach and progress with an instructor, a tutoring system, or their peers. They may also search for information on the Internet. Having the same assignment for all students in class is standard practice efficient for learning and developing skills. However, it is prone to cheating when used in a summative assessment such as graded homework, a mid-term test, or a final exam. Students can easily share and submit correct answers without completing the assignment. In this paper, we discuss methods for automatic problem generation for hands-on tasks completed in a computer lab environment. Using this approach, each student receives personalized tasks. We developed software for generating and submitting these personalized tasks and conducted a case study. The software was used for creating and grading a homework assignment in an introductory security course enrolled by 207 students. The software revealed seven cases of suspicious submissions, which may constitute cheating. In addition, students and instructors welcomed the personalized assignments. Instructors commented that this approach scales well for large classes. 
Students rarely encountered issues while running their personalized lab environment. 
Finally, we have released the open-source software to enable other educators to use it in their courses and learning environments.
\end{abstract}



\maketitle

\section{Introduction}

Mastering networking, operating systems, and cybersecurity is inconceivable without 
practical experience with real computer systems and tools. Practicing these skills in a physical or virtual laboratory or at students' own hosts is a common instructional practice. In general, students receive task assignments that are prepared to be solved in the provided lab environment (single or multiple connected computers). These 
assignments could be a part of on-premise or remote sessions with an instructor, individual or team homework, or extracurricular competitions. Regardless of the used instructional method, all students usually receive the same assignment and an instance of the same environment for its solving. They are expected to achieve the same goal and submit the same answer.

Having the same learning environment is convenient for learning new skills and technologies, but it is not always suitable for evaluating student learning or competitions. Students may share the correct answers with their peers 
via online communication. This answer sharing is easier than in other disciplines because the assignments themselves require interactions with computers. Monitoring students during the evaluation is laborious. 
Disconnecting students' computers from the Internet is impractical or infeasible because searching online sources such as documentation or data 
might be an inherent part of the assignment. Therefore, cheating is an issue that many computing educators face.

\textit{Automatic Problem Generation} (APG, also \textit{Automated Exercise Generation}) enables instructors to create modified versions of the problems (tasks), called problem instances. Each student is provided with one instance of the same problem. APG can thus mitigate the threat of copied or leaked answers~\cite{burket2015}. Although APG has already been applied in computing disciplines, it is not commonly used in hands-on assignments involving a lab environment (see \Cref{sec:related}).

This paper contributes to the broader adoption of APG by 
i)~providing an open-source toolset for an automated setup of a lab environment with unique configuration for each student (\Cref{sec:toolset},~\cite{apg-toolset}), and ii)~reporting experience from a case study of using the toolset in an introductory security course enrolled by 207 students (\Cref{sec:study}). The toolset enabled instructors to assign personalized hands-on homework to all students. The additional effort was minimal compared to the standard practice of assigning the same tasks to all students. In return, the toolset revealed seven groups of submissions indicating forbidden cooperation, such as sharing answers between students (\Cref{sec:results_discussion}). 


\section{Related Work} \label{sec:related}
%


Cheating and its mitigation is summarized in a recent paper~\cite[Section~2.3]{Fowler2021}. APG is one of the methods, which we focus on.


\subsection{APG in Lab Environment} \label{sec:apg_lab} 

This subsection reviews existing works on APG in hands-on assignments featuring a \emph{lab environment} (or \emph{sandbox}), which we define as one or more physical or virtual computers and/or networks the students interact with to solve the assignment. 

Security Scenario Generator (SecGen)~\cite{Schreuders2017} is a robust framework for building networks of virtual machines with randomized services, vulnerabilities, and themed content. SecGen can randomly choose elements of generated machines, such as operating systems, network 
services, user credentials, or vulnerabilities. Configuration of each element of the machines can be varied, such as network ports or the strength of passwords. SecGen uses its own complex scenario specification language in the XML format to describe constraints and properties of the generated machines, such as a system with a remotely exploitable vulnerability that would grant user-level access. 
SecGen has been used for teaching at universities and hosting a country-wide security competition in the United Kingdom. 
%

Chothia and Novakovic~\cite{Chothia2015} developed a virtual machine framework for cybersecurity education at the University of Birmingham, United Kingdom. The virtual machine runs a Linux operating system with many user accounts and intentionally flawed access control, web server, database server, and purposely-built insecure protocols. Once the machine is booted for the first time, its unique content is generated for each student. Students are tasked to find particular text strings (flags) and submit them to the server, which checks their correctness. The authors used the framework for exercises in introductory cybersecurity courses for master’s degree and undergraduate students. The exercises covered encryption, access control, key-agreement protocols, web security, and reverse engineering. 
The authors encountered three cases where groups of students copied the flags or shared the virtual machine. 

Tele-Lab~\cite{Willems2012} is an online lab environment with the automatic assessment of practical exercises in cybersecurity. The assignments can include variables, which are instantiated and used for creating personalized content of virtual machines in the lab. The variables are used in multiple-choice or free-text tests in a web interface of the lab environment. Students work with personalized machines and answer the tests with personalized content. The exercises cover attacks on accounts and passwords, network reconnaissance, eavesdropping of network traffic, wireless and web security. 

\subsection{APG Not Involving the Lab Environment} \label{sec:apg_nonlab} 
This subsection reports applications of APG 
where students do not interact with full-fledged computer hosts or networks.

Burket et al.~\cite{burket2015} deployed APG in 2014 in PicoCTF, a large-scale cybersecurity competition for middle- and high-school students. They used \emph{problem templates} to generate a pool of problem instances with unique answers per instance. The competition included ten automatically generated problems: five on cryptography, three featuring web pages, one on reverse engineering, and one on converting a number to a different base. 

Agudo et al.~\cite{Agudo2019} designed SERA, an extensible framework for personalized exercises for introductory computer security courses. The exercises are defined using their own specification language, which works with modules for generating assignments and checking students' submissions.
Each exercise is defined by i) the assignment template, 
ii) template parameters and their ranges used for the generation, and iii) functions for checking the students' submissions and providing feedback.
SERA has been piloted at the University of Malaga, Spain, using exercises on X.509 and TLS certificates, vulnerabilities of web servers, and secure e-mail.

MetaCTF~\cite{Feng2015} is a set of 17 homework assignments on reverse engineering and malware analysis for Linux operating systems. The assignments are organized into levels with increasing difficulty. Each level is completed when a student runs a provided binary, enters a correct password, and causes the binary to print 
the string \enquote{Good Job}. The binaries are unique for each student. MetaCTF includes a web interface for distributing individual binaries to students and checking the submitted passwords. MetaCTF was used in a course at Portland State University, USA, in 2015. 

Sadigh et al.~\cite{sadigh2012} reported their work-in-progress on applying APG for an undergraduate embedded systems course at the University of California, Berkeley, USA, 
in 2012.
They generated state-machines and real-time scheduling problems 
using problem templates and techniques 
for formal verification and synthesis. 

Fowler and Zilles~\cite{Fowler2021} focused on assessing basic programming skills. They created question variants of a similar difficulty using surface feature permutations. The variants were derived from base questions by changing specific elements such as variable or function names and the order of the parameters. The variant questions were used in homework assignments and the exam in an introductory Python course at the University of Illinois, USA, in 2020. 

Qi and Fossati~\cite{Qi2020} introduced Unlimited Trace Tutor, which automatically generates original blocks of Java code for practicing code tracing of \texttt{for} and \texttt{while} loops and \texttt{if} statements. The system generates a parse tree from a provided code snippet, modifies variables' values and relational operators, and produces a new snippet. 
The authors ran a pilot experiment with 11 volunteer students of Emory University, USA.

%

\subsection{Our Contribution} \label{sec:apg_our}

The contribution of this paper is in three areas. We enable fair summative assessment in the lab environment, report results of a study in the large class, and provide our toolset to other educators.

\subsubsection{Approach}

Our approach and technical solution appear similar to Tele-Lab~\cite{Willems2012}, which involves the lab environment and automatically assesses personalized assignments. However, no implementation or evaluation of Tele-Lab personalized assignments has been published after the initial paper from 2012.

Our approach is close to~\cite{Fowler2021}, which uses base questions and permutes their specific elements. However, \cite{Fowler2021} does not target networking, operating systems, or cybersecurity and does not involve any lab environment. Also, picoCTF~\cite{burket2015}, MetaCTF~\cite{Feng2015}, and SERA~\cite{Agudo2019} use templates for generating problems or files for cybersecurity competitions or classes. Still, they do not create the whole lab environment (virtual machines or networks), only their parts (web pages, binaries, certificates). 
PicoCTF and SERA provide a programming interface for generating arbitrary tasks and values. While this approach is more flexible than ours, instructors have to provide code for the generation instead of declaring type and constraints in the YAML markup language as in our approach.


Our technical solution is the most similar to~\cite{Chothia2015}.
While~\cite{Chothia2015} provides a single virtual machine that is difficult to modify, we enable educators to generate arbitrary networks with multiple machines. Next, unlike~\cite{Chothia2015}, which developed a custom submission server that is not publicly available, we extended CTFd~\cite{chung2017}, a popular open-source platform for hosting competitions and exercises. 

SecGen~\cite{Schreuders2017} applies APG in the lab environment with a different goal. Our goal is to provide students with a lab environment with the same structure but different content and values that students are required to search for. SecGen, on the other hand, creates various environments of the same complexity based on instructor-defined constraints. As a result, two environments generated by SecGen
can feature different network services and vulnerabilities.

Works~\cite{sadigh2012} and~\cite{Qi2020} do not involve the lab environment, and use approaches specific to the problem they generate.

\subsubsection{Evaluation}

This paper evaluates our method and toolset in the authentic teaching context of a large class. The same applies to~\cite{Fowler2021}, but they do not involve the lab environment. In contrast, APG by~\cite{burket2015} was evaluated in a different context (team security competition), though attended by 10,000 students in 3,000 teams. Similarly, SecGen~\cite{Schreuders2017} was used in another competition, but only with 59 students. Other works included evaluation with only a few participants (MetaCTF~\cite{Feng2015}, Unlimited Trace Tutor~\cite{Qi2020}), did not report the number of participants in their studies (\cite{Chothia2015}), or did not include any evaluation at all (Tele-Lab~\cite{Willems2012}, SERA~\cite{Agudo2019}, \cite{sadigh2012}).

\subsubsection{Reusability}

We have released our toolset as an open-source software project. Only the authors of SecGen~\cite{Schreuders2017} and picoCTF~\cite{burket2015} did the same, and partially~\cite{Chothia2015}. Unlimited Trace Tutor~\cite{Qi2020} is available free upon request. 

\section{Toolset for APG for Hands-on Labs} 
\label{sec:toolset}

\begin{figure}[t]
\centering
\includegraphics[width=0.95\linewidth]{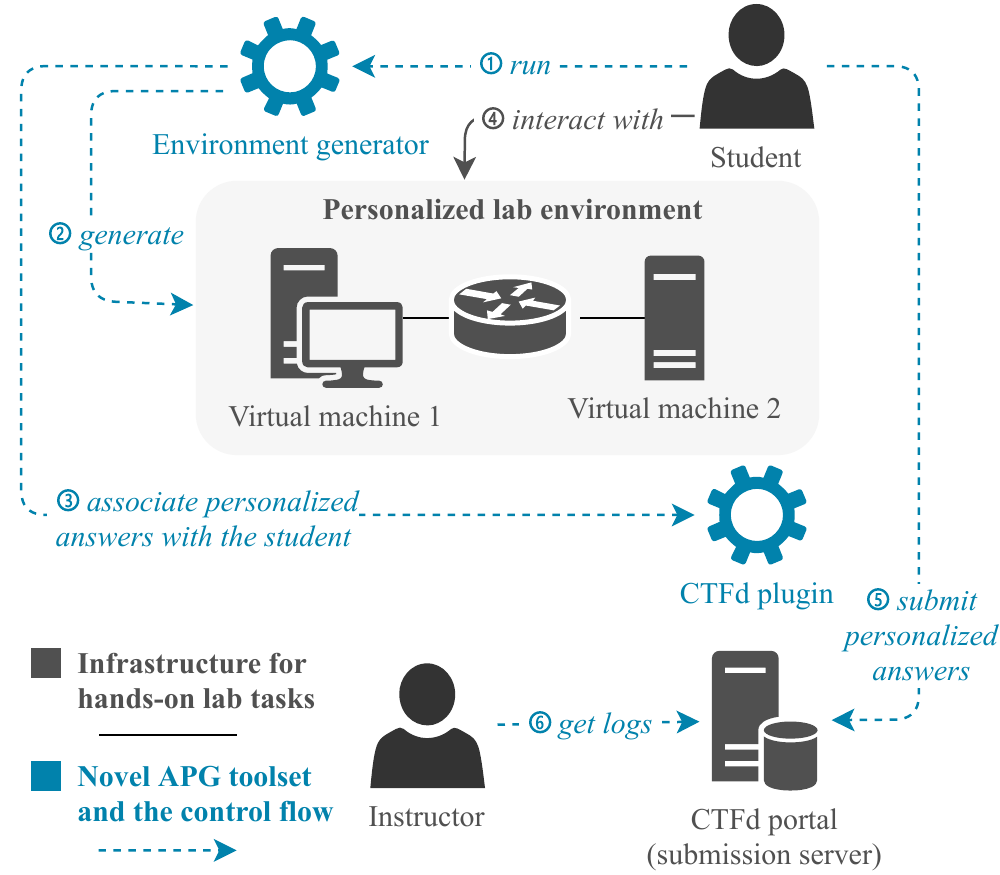}
\caption{The APG toolset design. \circlednumber{1} The student starts the environment generator with a unique seed. \circlednumber{2} The generator creates a personalized lab environment. \circlednumber{3} Answers specific to that environment are stored in the submission database. \circlednumber{4} The student solves the tasks. \circlednumber{5} The submitted answers are checked against the generated personalized answers. \circlednumber{6} Instructor examines the submission logs for cheating.} 
\label{fig:toolset}
\vspace{-0.4cm}
\end{figure}

Since no toolset for generating a personalized lab environment for summative assessment was available,  
we implemented it. The toolset consists of two core components, the \textit{environment generator} and the \textit{submission server}, see \Cref{fig:toolset}.

\subsection{Environment Generator}

Before developing the APG toolset, the lab environment where the students solved hands-on tasks was static, i.e., the same for everyone. The static lab environment was defined by text files specifying the parameters and configuration of virtual machines that students used for solving the hands-on assignments. Students were given these files and instantiated the lab environment locally on their computers using Cyber Sandbox Creator~\cite{my-2021-FIE-KYPO-CSC} based on free and open-source tools Vagrant~\cite{vagrant}, VirtualBox~\cite{virtualbox}, and Ansible~\cite{ansible}.

The new environment generator is a set of Python scripts that work on top of these lab definition files. A student instantiates their lab environment with a unique identifier, such as an e-mail address. This identifier is hashed and serves as a seed for generating values used to configure and deploy the personalized lab environment. For example, network service will run on a different unique port for each student in a networking lab.

The environment generation is driven by a configuration file that specifies the types of the generated values and their constraints, such as a range or prohibited values. Currently, it can generate user names and passwords of a predefined length, sentences, IP addresses, and port numbers. The generated random values are passed to Ansible, which is responsible for configuration management of the virtual machines in the environment. 

\subsection{Submission Server}

Students who solve hands-on tasks enter their answers into a central submission server. However, in traditional static assignments, every student searches for the same answer, which is prone to cheating.

Along with the environment generator, we developed an open-source plugin that extends the popular platform CTFd~\cite{chung2017}. CTFd is a web portal that allows students to submit answers to instructor-defined tasks. When the students instantiate their personalized lab environment, the environment generator sends the generated personalized answers to CTFd. Our plugin then associates individual answers with the corresponding students in the database. 

This submission server brings two key improvements over the traditional assessment process. First, for each student, only their personalized answer is accepted as correct. Second, if a student submits an answer of someone else, this activity is logged.

\section{Case Study Methods} \label{sec:study}

We describe the methods of the case study that uses the proposed toolset for APG. The goal is to evaluate APG for hands-on lab assignments in an authentic teaching context.

\subsection{Teaching Context and Participants}

The case study involves one homework assignment in an introductory computer security course enrolled by 207 undergraduate students. The course was taught at a public university in the Czech Republic in the Spring 2021 semester remotely (via video conferences) in English. 
The homework constituted 4\% of the final grade and was due in 14 days (May 3--17, 2021).

\subsection{Exercise Content and Format}
The assignment enhanced the skills students learned in the lab session before homework. In particular, it covered network attacks on authentication of Telnet and SSH servers, securing an SSH server, and capturing and analyzing SSH traffic. 
The assignment had to be completed individually using the personalized lab environment and submission server introduced in \Cref{sec:toolset}. 

\subsubsection{Task personalization} The homework was structured into eight tasks. Each task had to be completed by entering a text string (answer) to the submission server. Answers to three tasks were uniform, and five tasks were personalized for each student. These answers were generated using our APG toolset. 
In the end, each student had a personalized lab environment, which contained i) a host running the Telnet server at a random network port, ii) one user account with a random username, iii) another user account with a random password, and iv) a file containing a random sentence. The generation of the random port number was limited to the values from 1500 to 65000, excluding the number 2323, a well-known alternative port number for the Telnet service. The password was a four-digit number from 1300 to 2000, excluding the numbers 1234 and 1337 because some students may guess these numbers. Finally, the username was one of the common usernames from a well-known dictionary used for authentication attacks.






\subsubsection{Task dependencies} 
\begin{figure}[t]
    \centering
    \begin{tikzpicture}[
  ->,                  
  >=stealth',          
  shorten >=1pt,    
  auto,
  node distance=1.3cm,  
  thick,
  scale=0.8,
  every node/.style={scale=0.7} %
  ]
  \node[state,color=black,initial]   (F)            {F};
  \node[state,color=ACMBlue] (A1) [above right of=F]   {A1};
  \node[state,color=gray,style=dashed] (A2) [right of=A1]        {A2};
  \node[state,color=gray,style=dashed] (A3) [right of=A2]        {A3};
  \node[state,color=gray,style=dashed] (A4) [right of=A3]        {A4};
  \node[state,color=gray,style=dashed] (A5) [right of=A4]        {S1};
  \node[state,color=gray,style=dashed] (A6) [right of=A5]        {S2};
  \node[state,color=ACMBlue] (T1) [right of=F]         {T1};
  \node[state,color=ACMBlue] (T2) [below right of=F]   {T2};

  \path (F) edge node {} (A1)
        (A1) edge node {} (A2)
        (A2) edge node {} (A3)
        (A3) edge node {} (A4)
        (A4) edge node {} (A5)
        (A5) edge node {} (A6)
        (F) edge node {} (T1)
        (F) edge node {} (T2)
        ;
    
    \end{tikzpicture}
    \caption{Blue tasks are displayed after a student answers the familiarization question (F). Locked (hidden) tasks are gray.}
    \label{fig:chal_dep}
\end{figure}
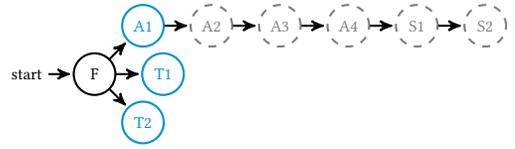

Before a student started to solve the homework tasks, they were presented with one simple question, intended only for their familiarization with the submission server and conventions. The familiarization explained submitting the answers, displaying hints, and unlocking tasks in a chain. Once they solved it, they were allowed to start solving the homework tasks. In the beginning, they could choose the first task from a chain of six tasks (A1--A4, S1, and S2), or any of the two other tasks (T1 or T2), as depicted in~\Cref{fig:chal_dep}. The chain defined the order of the six tasks, unlocked once a student had solved the previous tasks in the chain. Each task was worth 1 point. 

\subsubsection{Homework rules} \label{sec:rules}
Students were instructed that sharing correct answers is strictly prohibited and useless because each of them had received the personalized lab environment.
Next, students had five attempts to submit a correct answer to each task. After that, they were not allowed to submit another answer or proceed to the next task in a chain. Finally, when the homework was over, we randomly selected four students who were required to demonstrate their approach to the instructor at a dedicated one-to-one video call. All these rules were announced when the homework was assigned.

\subsection{Cheating Detection} 

To reach the goal of this study, we also detect suspicious students' submissions, which may indicate cheating. Some of the methods were piloted in our previous research~\cite{vykopal2020-ctf}.

\subsubsection{Someone else's answers} The most reliable detection method is tracking incorrect submissions of correct answers belonging to other students. This method assumes that some students shared their correct answers with other students who unthinkingly submitted someone else's answers. 
\subsubsection{Task chains} Another method benefits from locked tasks in a chain. Since a task in the chain is unlocked only after the previous task is successfully solved, this method computes the solve time for consecutive tasks. Then, the student's solve time is compared to the \emph{minimal possible solve time} of a human who immediately performed all actions required to solve the tasks without any mistakes and time for thinking about the steps. Any student's solve time close to or lower than the minimal possible solve time may indicate that the student obtained step-by-step instructions from another student.

\subsubsection{Submission proximity} The least conclusive method lies in searching for \emph{time proximity} or \emph{location proximity} of two or more submissions. Any of these proximities may indicate that students were working together and submitted the answers (correct or incorrect) at the same time or place.
We consider submissions to be in the location proximity if they originated from the same IP address, which multiple hosts might share in some networks~\cite{Maier2011}.

\subsection{Data Collection and Survey}
To use the methods for cheating detection, we logged 
students' correct and incorrect answers submitted to the submission server, together with timestamps and IP addresses.

In addition, we surveyed students about their opinions on the format of the homework assignment. While the assignment was an inherent part of the course, the survey after the assignment was optional. All students who participated in the survey provided their informed consent to use the collected data for research purposes. \Cref{tab:postGameAssessmentQuestions} lists the questions asked in the survey. 

\begin{table}[h]
    \centering
    \caption{Wording of the survey questions.}
    \label{tab:postGameAssessmentQuestions}
    \small
    \begin{tabular}{p{0.3cm}|p{7.3cm}}
    \toprule 
    \textbf{No.} & \textbf{Question} \\  \midrule
        Q1 & Would you rather complete an assignment of this format (lab environment~+~submission server) or traditional homework assignments (written report) in your future security courses?  \\ \midrule
        Q2 & How useful was the instant feedback on your submissions (the server's response whether your answer is correct or not)?  \\ \midrule
        Q3 & How stressful was the limited number of five attempts per task?  \\ \midrule
        Q4 & How stressful was the possibility that you could be randomly selected to demonstrate your solution approach to the instructor?  \\ \midrule
        Q5 & Have you experienced any technical issues with your personalized lab environment?  \\ \midrule
        Q6 & Do you have any comments or thoughts related to the previous questions or any other feedback?  \\ 
    \bottomrule
    \end{tabular}
\end{table}

\section{Results and Discussion} \label{sec:results_discussion}

Here we report the study results and summarize our experience with APG and preventing cheating in lab assignments. Out of 207 students enrolled, 195 students logged into the submission server. In total, 178 students solved at least one task. All eight tasks were solved by 160 students. 


\subsection{Suspicious Submissions}

We analyzed the students' answers to identify suspicious submissions, which may indicate cheating.

\subsubsection{Someone else's answers}

We discovered three cases. 

\paragraph{Case 1} The most conclusive was the following case. Student A submitted the correct answer \texttt{41247} for A1 on May 9th, 22:39. Student B submitted the incorrect flag \texttt{41247} twice, several days later: first on May 13, 23:23, and then on May 14, 11:21. However, Student B generated his personalized lab environment for the first time on May 14, 11:00. That means the first incorrect submission of Student B occurred before his first interaction with the lab environment. 
When questioned, the students replied that Student B used a laptop of Student A with the already running personalized lab environment of Student A due to technical issues of a laptop of Student B. 
To avoid similar situations, instructors should communicate to students what constitutes cheating together with concrete examples of violations of homework rules.

\paragraph{Case 2} A1 involved using the \texttt{nmap} tool to discover a network service running on a personalized network port. Student C submitted an incorrect answer \texttt{16278} on May 11, 16:16:55. Student D submitted his correct answer \texttt{16278} only 90 seconds later. 
Then, Student C submitted his correct answer \texttt{26569} 4 minutes after Student D. 
We asked Student C why he had first submitted the incorrect answer \texttt{16278}. He replied he \enquote{typed a random number}. A more likely explanation is that Student D solved the task first and shared his correct answer with Student C who submitted it first. The submission server replied that the answer was incorrect for Student C. He then asked Student D for the command to run in his environment and later submitted the correct answer. The network service discovery lasted only a few seconds, so Student C could have completed the task in 4 minutes. 

\paragraph{Case 3}
Student E submitted the correct answer \texttt{asd} for A4 on May 8, 11:39. Student F submitted an incorrect answer \texttt{asd} for the familiarization question six days later. 
Since one of the tasks involved was the familiarization question and \texttt{asd} is a common testing string composed from three neighboring letters on a keyboard, we do not consider this case cheating.

\subsubsection{Task chains} 


We discovered two cases where students submitted their answers incredibly quickly. 

\paragraph{Case 4} The minimal possible solve time of A3 was 45 seconds. The assignment text consists of 102 words. The task requires capturing the network traffic using the \texttt{tshark} tool to intercept the secret content of the file. The tool has to be configured to analyze packets at a non-standard network port discovered in A1.
Three students G, H, and I completed the task in 58 seconds.
These solve times contrast with other tasks, which they completed in near to median time compared to the other students. We consider their submissions suspicious since reading the assignment text, thinking, and performing all the actions take considerably longer than the minimal possible solve. However, the students may not have followed the assignment text (i.e., have not intercepted the network traffic) but simply printed out the file content as another student confessed during one demonstration session (see \Cref{sec:results_demonstration}).

\paragraph{Case 5} The minimal possible solve time of S2 was 33 seconds. The assignment text of S2 consists of 69 words. The task requires disabling public key authentication at an SSH server. This involves setting the right configuration option and restarting the server. Student J completed the task in 46 seconds. Since the task involved multiple actions in the lab environment, it is unlikely to accomplish them in such a short time. Since the answer to this task was a short text string same for all students, the student may have obtained it from another student or found the answer in the documentation without applying it in the environment. One option for personalizing this task would be to insert a commented string into the configuration file on a random line and ask for a line number that must be uncommented.


\subsubsection{Submission proximity}

We found one confirmed case of students' collaboration using the location proximity and the least conclusive case using time proximity.

\paragraph{Case 6} 
Four groups of students used the same IP address. Students from three groups submitted their correct answers at different times. However, two students K and L in one group submitted their answers to T2 within 68 seconds. Student K confessed he had cooperated with student L. He told us they share the same dormitory room and shared only the steps for T2, not the answer.

\paragraph{Case 7} 
Assuming the homework was open for 14 days, and the students worked individually, it is improbable that they submitted the correct answer at the almost same time. We found that students M and N submitted their answers for A3 within 13 minutes, for A4 within 2 minutes, for S1 within 13 minutes, and for S2 within 4 minutes (all at midnight of May 16). They submitted the other tasks within one hour. Still, it might be a coincidence since they worked the last day before the deadline, so there was a higher chance of submissions in close time proximity.

\subsection{Findings from Demonstration Sessions} \label{sec:results_demonstration}

One of the demonstration sessions after the homework deadline revealed that one student did not complete A3 as required. He only simply printed out the content of a file at the server instead of capturing the network traffic using the \texttt{tshark} tool. The point he had earned for submitting the correct answer for this particular task was deducted. The other three sessions did not reveal any issues.


\subsection{Results from the Post-Homework Survey}
The optional survey after the assignment was answered by 45 students. 
Forty students (89\%) reported they would prefer the provided format of completing assignments in security courses. Only one student would prefer the traditional homework assignment, and the remaining four were not sure. 

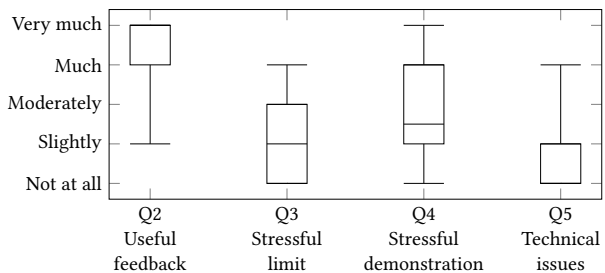
\begin{figure}[h!]
    \centering
    \small
    \scalebox{0.95}{
        \begin{tikzpicture}
            \begin{axis} [
                xtick={1,2,3,4},
                xticklabel style={align=center},
                xticklabels={Q2\\Useful\\feedback, Q3\\Stressful\\limit, Q4\\Stressful\\demonstration, Q5\\Technical\\issues},
                ytick={1,2,3,4,5},
                yticklabels={Not at all, Slightly, Moderately, Much, Very much},
                width=1\linewidth,
                height=0.5\linewidth
            ]
                \addplot [box plot median] table [opacity=0.5, color=orange]{boxplotdatasurvey.dat};
                \addplot [box plot box] table {boxplotdatasurvey.dat};
                \addplot [box plot top whisker] table {boxplotdatasurvey.dat};
                \addplot [box plot bottom whisker] table {boxplotdatasurvey.dat};
            \end{axis}
        \end{tikzpicture}
    }
    \caption{Answers to Q2--Q5 in the survey from 45 students.}
    \label{fig:answers-Q2-Q5}
\end{figure}

The answers to questions Q2--Q5 are summarized in \Cref{fig:answers-Q2-Q5}.
The students highlighted that the immediate feedback of correctness or incorrectness of their answers was much useful (Q2). Next, most students considered the limited number of attempts to answer the tasks to be stressful slightly or not at all (Q3). Answers to Q4 show the students were stressed a bit more about the possibility to be chosen for the solution demonstration. This may indicate (i) students attempted cheating, or (ii) students were not sure whether the instructor would approve of their approach (not the answer). Answers to Q5 show that the APG toolset worked well. 

Open-ended Q6 yielded diverse answers. 
Nine students elaborated their answer to Q1 that they enjoyed the homework format. 
Three students rated this homework as one of the best assignments in the course.
Four students rated the difficulty of the particular tasks and sometimes suggested improvements. Two students mentioned not having sufficient system resources for a smooth run of the lab environment. Two other students reported other technical issues with running the environment at their own hosts. Two students elaborated answers to Q3 about the submission limit. One student \enquote{wasn't really worried about limited attempts because it was clear what we are supposed to submit}. Another student reported: \enquote{At first, I thought those 5 attempts will be very limiting, but after solving it, I realized, it wasn't limiting at all}.



\subsection{Limitations}

Our study involved a single exercise in one course. Still, the number of participants who interacted with our tool is considerably larger than in the vast majority of works reported in \Cref{sec:related}.

The cheating detection methods analyze only students' actions at the submission server. We do not capture any other data, such as commands typed in the lab environment. We can confront students with our findings, but we cannot be entirely sure whether students actually cheated or not (see Case 1).

Estimating the location proximity using the same IP address of the submission is a double-edged sword. While our study shows that one IP address may be shared by roommates, in other cases it might be shared by students who work independently.

Since the personalized answers were generated at students' computers, advanced students may reverse-engineer the environment generator and obtain the answers even without interaction with the personalized lab environment. However, we agree with~\cite{Chothia2015} that doing so would be more difficult than completing the assignment.

The optional survey was answered by 45 students out of 195 who started solving the homework assignment. The answers may not entirely represent the opinions of all students solving the homework, particularly the critical voices. Nevertheless, we have not received any negative feedback from other formal or informal channels.

\section{Conclusions and Future Work} \label{sec:conclusion}

We presented an open-source toolset for creating and marking personalized hands-on assignments involving virtual machines and networks. The toolset was used for preventing and detecting cheating in individual homework in an introductory security course enrolled by 207 students. Each student received tasks with the same assignment text but different answers to be found by interacting with the lab environment. We discovered seven suspicious cases using three different cheating detection methods. At the same time, students enjoyed the assignment and its format and did not perceive cheating prevention disruptively. 

Our approach is lightweight and privacy-preserving. Students were not under surveillance when solving their homework. Still, we discovered suspicious submissions only from minimal data collected (submitted answers, timestamps, IP addresses). Logging additional student actions 
would increase the precision of cheating detection. We plan to integrate a command-line logging toolset~\cite{my-2021-FIE-logging} 
into the personalized lab environment. The ability to analyze the students' commands would allow to reconstruct the problem-solving process and timeline better.

To conclude, we showed that prevention and detection of cheating in hands-on assignments involving the lab environment is possible in large and remote classes. The key components are: automated provisioning of the lab environment with personalized values generated locally at students' computers, task chains, submission limits, and a demonstration session. 
We provide the toolset and an exemplary assignment in a public repository~\cite{apg-toolset}. Since all used components are free and open-source, other instructors can immediately use them in their classes as we did or adapt only the components that fit their needs. 

Our future work will focus on automatic problem generation preserving tasks' difficulty and fidelity of the lab environment for teaching networking, operating systems, and cybersecurity skills. An example of a challenging project is the automatic generation of a network with several hosts, each with a random yet valid IP address, able to communicate with other hosts in the network.

\begin{acks}
This research was supported by \grantsponsor{ERDF}{ERDF}{} project \textit{CyberSecurity, CyberCrime and Critical Information Infrastructures Center of Excellence} (No. \grantnum{ERDF}{CZ.02.1.01/0.0/0.0/16\_019/0000822}). We also thank Daniel Košč for developing the toolset.
\end{acks}

\balance
\bibliographystyle{ACM-Reference-Format}
\bibliography{references}


\end{document}